\documentclass[11pt,fleqn,a4]{article}
\usepackage{graphicx} 
\usepackage{amsmath}
\usepackage{booktabs} 
\usepackage{multirow} 
\usepackage{txfonts}
\usepackage{bm}
\usepackage{color} 
\usepackage{soul} 
\usepackage{graphicx}
\usepackage{caption}
\usepackage{subcaption}

\makeatletter
\def\ps@isopefoot{\let\@mkboth=\@gobbletwo
 \def\@oddhead{}
 \def\@evenhead{}
 \def\@oddfoot{\hbox to \textwidth
   {\footnotesize {Paper No.\,2011-\papernumber} \hfil
   {\firstauthorname} \hfil {Page number:\ \thepage\ of\ \totalpage}}}
 \def\@evenfoot{\hbox to \textwidth
   {\footnotesize {Paper No.\,2011-\papernumber} \hfil
   {\firstauthorname} \hfil {Page number:\ \thepage\ of\ \totalpage}}}
}%
\def\leftcapmargin{35pt}
\def\rightcapmargin{15pt}
\long\def\@makecaption#1#2{%
 \vskip 10\p@ \footnotesize\baselineskip=4.0mm
 \@tempdimb=\leftcapmargin
 \@tempdima\hsize\advance\@tempdima-\@tempdimb
 \@tempdimb=\rightcapmargin
 \advance\@tempdima-\@tempdimb
 \newbox\tempbox
 \setbox\tempbox=\hbox{#1}
 \setbox\@tempboxa=\hbox{\hskip\leftcapmargin #2\hskip\rightcapmargin}
 \ifdim \wd\@tempboxa >\hsize
 \hbox to\hsize{\hfil \hskip-\the\wd\tempbox \hskip\leftcapmargin #1
 \parbox[t]\@tempdima{#2}\hskip\rightcapmargin}%
 \else \hbox to\hsize{\hfil #1 #2\hfil}\fi
}
\def\fnum@figure{{\footnotesize Fig.\,\thefigure}\ }
\def\fnum@table{{\footnotesize Table\,\thetable}\,}
\def\section{\@startsection{section}{1}{\z@}
{2ex plus .3ex minus .2ex}{1.5ex plus .1ex}{\small}}
\def\subsection{\@startsection{subsection}{2}{\z@}
{1.5ex plus .2ex minus .2ex}{1ex plus .1ex}{\small\bf}}
\def\subsubsection{\@startsection{subsubsection}{3}{\z@}
{1.5ex plus .2ex minus .2ex}{1ex plus .1ex}{\footnotesize\it}}
\def\paragraph{\@startsection{paragraph}{4}{\z@}
{1ex plus .3ex minus .2ex}{-1em}{\large\bf}}
\def\subparagraph{\@startsection{subparagraph}{4}
{\parindent}{1ex plus .3ex minus .2ex}{-1em}{\small\bf}}
\def\eqnarray{%
 \stepcounter{equation}%
 \let\@currentlabel=\theequation
 \global\@eqnswtrue
 \global\@eqcnt\z@
 \tabskip\mathindent
 \let\\=\@eqncr
 $$\halign to \displaywidth\bgroup\@eqnsel\hskip\@centering
 $\displaystyle\tabskip\z@{##}$&\global\@eqcnt\@ne
 \hfil$\displaystyle{{}##{}}$\hfil
 &\global\@eqcnt\tw@$\displaystyle\tabskip\z@{##}$\hfil
 \tabskip\@centering&\llap{##}\tabskip\z@\cr}
\def\@cite#1#2{\,[{\hbox{#1\if@tempswa , #2\fi}}]}
\def\thebibliography#1{%
\begin{flushleft}\section*{REFERENCES}\end{flushleft}
\list
{[\arabic{enumi}]}{\settowidth\labelwidth{#1)}\leftmargin\labelwidth
 \advance\leftmargin\labelsep
 \usecounter{enumi}}
 \def\newblock{\hskip .11em plus .33em minus .07em}
 \sloppy
 \sfcode`\.=1000\relax}

\newif\if@lastpagecolumnalign \@lastpagecolumnaligntrue
\if@lastpagecolumnalign 
\newdimen\lastp@geheight \lastp@geheight=10mm
\newsavebox{\lastp@gebox}
\long\def\endabstract#1{\gdef\@endabstract{#1}} 
\def\lastpagecontrol{\@ifnextchar [{\l@stpagecontrol}%
 {\l@stpagecontrol[\z@]}}
\def\l@stpagecontrol[#1]#2{\global\lastp@geheight=#2%
 \@ifundefined{maxsize}{}{\global\advance\maxsize-#2}
 \@ifundefined{@endabstract}{}{
 \global\sbox{\lastp@gebox}{%
  \begin{minipage}{\textwidth}\vspace*{#1}%
  \hrule width \textwidth \vspace{1ex}%
  \begin{list}{}{\setlength{\leftmargin}{0.15\textwidth}%
  \listparindent=10pt \parsep=0pt
  \topsep=0pt \partopsep=0pt
  \setlength{\rightmargin}{\leftmargin}\small}%
  \item \ignorespaces\hspace*{\listparindent}\ignorespaces
                 \@endabstract
  \end{list}%
  \vspace{1ex} \hrule width \textwidth
  \end{minipage}}}%
  \@tempdima\ht\lastp@gebox \advance\@tempdima\dp\lastp@gebox
 \ifdim\@tempdima>\lastp@geheight
  \@tempdima\lastp@geheight \global\lastp@geheight=0pt
 \else
  \global\advance\lastp@geheight -\@tempdima
  \@tempdima\lastp@geheight \global\lastp@geheight\textheight
 \fi
  \def\footnoterule{\null}
  \insert\footins{\footnotesize
  \interlinepenalty\interfootnotelinepenalty
  \splittopskip\footnotesep
  \splitmaxdepth \dp\strutbox \floatingpenalty \@MM
  \hsize\textwidth \@parboxrestore
  \ifdim\lastp@geheight=\z@\else\usebox{\lastp@gebox}\fi%
  \vspace*{\@tempdima}}}
\def\lastpagesettings{\@ifnextchar [{\l@stpagesettings}%
 {\l@stpagesettings[\z@]}}
\def\l@stpagesettings[#1]{%
 \ifdim\lastp@geheight=\z@
 \onecolumn\null\vspace*{#1}\noindent\usebox{\lastp@gebox}%
 \fi}
\fi 
\makeatother
\def\jsection#1{\section{\hspace*{-4.0mm}\uppercase{#1}}}

\newcommand{\bdm}{\begin{displaymath}}
\newcommand{\edm}{\end{displaymath}}
\newcommand{\be}{\begin{equation}}
\newcommand{\ee}{\end{equation}}
\newcommand{\bea}{\begin{eqnarray}}
\newcommand{\eea}{\end{eqnarray}}
\newcommand{\bc}{\begin{center}}
\newcommand{\ec}{\end{center}}
\newcommand{\bfl}{\begin{flushleft}}
\newcommand{\efl}{\end{flushleft}}
\newcommand{\bfr}{\begin{flushright}}
\newcommand{\efr}{\end{flushright}}

\newlength{\mybaselineskip}
\newlength{\myleftmargin}
\newlength{\mytopmargin}
\renewcommand{\vec}[1]{\mathbf{#1}}
\newcommand{\papernumber} {TPC-0296}
\newcommand{\firstauthorname} {Juhan Wang}
\newcommand{\totalpage} {7}
\begin{document}
\baselineskip \mybaselineskip \twocolumn[
\par\vspace*{30.0mm}\large\bc{\bf
%
An active learning strategy to study the flow control of a stationary cylinder with two asymmetrically attached rotating cylinders\\
}\ec %
\par\vspace{1mm}\footnotesize\bc{\small\it
Juhan Wang (Master Student)
}\\ 
Naval Architecture and Marine Engineering, University of Michigan\\
Ann Arbor, Michigan, USA 

\par\vspace{3mm}
{\small\it
Dixia Fan
}\\ 
Mechanical Engineering, Massachusetts Institute of Technology\\
Cambridge, Massachusetts, USA

%
\ec\par\vspace*{22mm}
]
\footnotesize\par\noindent{\small ABSTRACT}\par\vspace*{2.0mm}
\baselineskip \mybaselineskip
\setstcolor{red} 
We numerically investigate the flow control problem of the flow passing a stationary cylinder at a fixed Reynold number 500 using two attached control cylinders with different rotation rates. Compared to the traditional uniform (lattice) sampling method, we developed an active learning strategy based on Gaussian Process Regression (GPR), drastically reducing the number of simulations and accelerating the scientific findings. We also discussed the effects of parameters on different hydrodynamic coefficients, and verified the feasibility of this strategy. The mechanism of this asymmetric flow control model was also further studied by analyzing flow patterns.
\footnotesize
\par\bigskip\noindent{{\small KEY\,\,WORDS}}:
Active Learning; Flow Control; Numerical Simulation;
\par\vspace*{2mm}%
\jsection{Introduction}
Fatigues and vibrations due to dynamic marine environment, like Vortex-Induced Vibration(VIV), are fairly severe that the analysis of fluid and structure interaction (FSI) for offshore structures are important. Since almost all of these structures have crucial components such as piles and stacks that could be modeled as cylinders in fluid domain, we narrow our focus on the flow around the cylinder, one representative type of bluff body, where there exists well-known studies having figured out the mechanism of oscillations (Sarpkaya, 1979) and the crucial effect of Reynold number (Pastò, 2008) of this FSI problem, as well as different wake modes of two side-by-side circular cylinders at low Reynolds numbers (Sangmo, 2003) and fluid characteristics of rotating cylinders next to a wall (Cheng et al., 2007; Rao et al., 2011). Almost all these studies aim at how to enhance the hydrodynamic properties of target bodies, promoting a vivid flow control field consisting of various active (Jahanmiri, 2010) and passive control strategies (Kumar et al., 2008), with related studies (Sakamoto et al., 1994) on their mechanism and the nature of the controlled wake. However, though extra energy is needed, active flow control, with Moving Surface Boundary-Layer Control (Modi, 1997) as the most frequently used method, has been regarded to be more robust and flexible than its counterpart. One model incorporating this methodology is to use two attached rotating cylinders (Muddada et al., 2010), which is also the prototype of our improved model. Specifically, With an objective of drag reduction and focusing on the mean drag coefficient ($C_D^{MEAN}$), this model could achieve higher energy efficiency of underwater vehicles, while their maneuverability could be improved with the mean lift coefficient ($C_L^{MEAN}$) studied. Similarly, When this method used in wave energy generation with Root-Mean-Square (RMS) values of the drag and the lift force ($C_D^{RMS}$, $C_L^{RMS}$) targeted, the productivity of energy harvest device could be improved by magnifying vibration amplitudes. \\

Previous studies have figured out a few effects of some parameters in this model. For example, the lift-to-drag ratio could be obviously improved with small control cylinders rotating at a rate higher than one, normalized by the upcoming fluid velocity (Modi, 1997), while the flow separation of the target main cylinder could be delayed even with two stationary small cylinders (Kuo et al., 2007). Besides, with two symmetrically attached counter-rotating cylinders at same rates, the critical range of the normalised gap value ($g/D$, where $g$ is the gap from small control cylinders to the main cylinder whose diameter is $D$) was found (Mittal, 2001, 2003), the mechanism behind which is its key effects on the length of recirculation zone, thereby affecting the hydrodynamics of the target body. From this analysis, the normalized gap in our study is determined as 0.1, neither too narrow to place sufficient grids nor too wide to deteriorate the performance of actuators. Furthermore, the control mechanism of this model is found to be viscous (Schulmeister et al., 2017), which means the crux of achieving flow control is to delay the separation point, which basically locates around 90$^{\circ}$ with respect to the downward stream at Reynold number 500. Therefore, this relative angle ($\theta$) in our study was selected from 30$^{\circ}$ to 120$^{\circ}$. Apart from the relative location of control cylinders, the proportional relationship between the diameter ratio ($d/D$, diameter of small cylinders over that of the main cylinder) and the control effects is also a crucial characteristic of this model (Schulmeister et al., 2017), paving roads for us to select it as 0.125 in our study. As for the rotation rates ($\gamma$), normalized by the velocity of upcoming flow with positive value representing rotating clockwise, vice versa, the norm limit in our study was determined as 5, for the control system requires higher energy input but makes smaller contribution rates greater than 5 (Choi et al., 2008). \\

However, this pervasively used symmetrical model discussed above has some constraints in practical applications and is limited to find more effective layouts to achieve better control effects. Hence, we reconfigured it and introduced asymmetry by using control cylinders rotating at different rates. Specifically, we used three parameters in our analysis: the relative angle ($\theta$) with respect to the downward stream and different rotation rates ($\gamma_1$, $\gamma_2$) of the upper and the lower control cylinders. What should be highlighted after the parametric design is that we adopted an active learning strategy based on Gaussian Process Regression (GPR) in our simulations instead of the traditional uniform (lattice) sampling method, drastically reducing the number of simulations and accelerating the scientific findings without loss of fidelity. \\

In the following sections, we will firstly conduct a numerical verification via comparing with existing studies on the hydrodynamics of an isolated cylinder, after which we will introduce our active learning algorithm. Then simulation results and the learning process will be demonstrated, and the mechanism of parametric effects on the hydrodynamics of the main cylinder will be discussed with flow patterns analyzed.\\

\medskip
\jsection{MODEL SETUP AND VERIFICATION}\par\vspace*{2mm}%
With the hydrodynamic force component along the flow direction determined as the drag force (Equation \ref{eq:C_D}) and that perpendicular to the flow direction (Equation \ref{eq:C_L}) as the lift force, four hydrodynamic coefficients ($C_D^{MEAN}$, $C_L^{MEAN}$, $C_D^{RMS}$, $C_L^{RMS}$) were selected in our study to qualify the hydrodynamic characteristics of the main cylinder with the mean values for vibration amplitudes and the RMS values for vibration intensity.
\begin{eqnarray}
  && C_D = \frac{F_x}{0.5\rho U_\infty^2 D}  \label{eq:C_D}\\
  && C_L =  \frac{F_y}{0.5\rho U_\infty^2 D} \label{eq:C_L}
\end{eqnarray}
where, $U_\infty$ is the velocity of upcoming flow.\\

Besides, the fluid turbulence is captured by Reynold Number (Re = $U_\infty D/\nu$), which was fixed at 500 in our study to avoid three-dimensional effects in the wake of target body, while the frequency of vortex shedding is described by Strouhal Number (St = $f D/U_\infty$), which was calculated via Fourier Transform in MATLAB.\\

\begin{figure}[htb]
\centering
\includegraphics[width=0.6\linewidth]{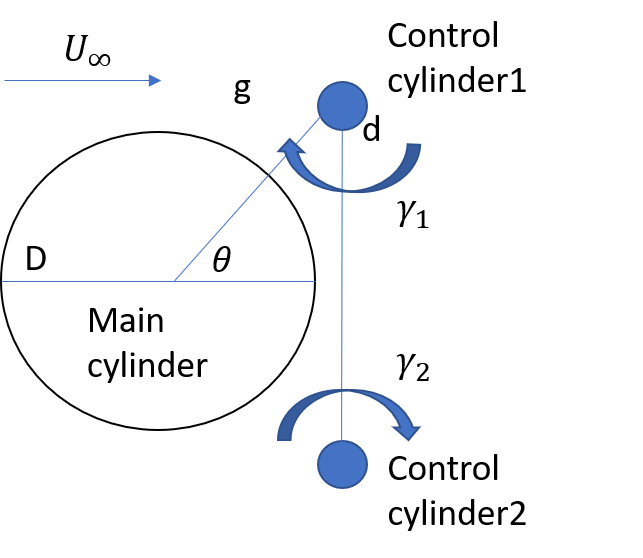}
\caption{Diagram of flow control model} \label{Fig:model}
\end{figure}

Also as introduced before, We framed our model (Figure \ref{Fig:model}) asymmetrically by setting the normalized rotation rates of the upper and the lower control cylinders different (Table \ref{tab:para}).\\

\begin{table}[htbp]\footnotesize
\vspace*{-4mm}
\begin{tabular}{p{30pt}p{30pt}p{30pt}p{30pt}p{40pt}p{30pt}}
    \toprule
Parameter & Re  & $g/D$ & $d/D$   & $\theta$  & $\gamma_1$, $\gamma_2$ \\
    \midrule
Value     & 500 & 0.1 & 0.125 & 30$^{\circ}$ $\sim$ 120$^{\circ}$ & -5 $\sim$ 5 \\
    \bottomrule
\end{tabular}
 \caption{Selection of parametric values} \label{tab:para}
\end{table}

After model construction, the simulation of an isolated cylinder without attached control cylinders was conducted with comparison with existing studies in terms of $C_D^{MEAN}$ and the Strouhal Number (Table \ref{tab:cdmean}, \ref{tab:st}).\\

\begin{table}[htbp]\footnotesize
\vspace*{-4mm}
\begin{tabular}{p{25pt}p{40pt}p{40pt}p{30pt}p{40pt}p{15pt}}
    \toprule
       & Schulmeister et al., 2017 & Henderson et al., 1995 & Ji et al., 2019 & Zhao et al., 2013 & We \\
    \midrule
    $C_D^{MEAN}$ & 1.326                     & 1.274                  & 1.363           & 1.225             & 1.328 \\
    \bottomrule
\end{tabular}
 \caption{Comparison of $C_D^{MEAN}$} \label{tab:cdmean}
\end{table}
\begin{table}[htbp]\footnotesize
\vspace*{-4mm}
\begin{tabular}{p{15pt}p{40pt}p{40pt}p{40pt}p{40pt}p{15pt}}
    \toprule
 & Ji et al., 2019 & Zhao et al., 2013 & Jiang et al., 2017 & Sirisup et al., 2004 & We \\
    \midrule
St & 0.211 & 0.208 & 0.206 & 0.22 & 0.24 \\
    \bottomrule
\end{tabular}
 \caption{Comparison of Strouhal Number} \label{tab:st}
\end{table}
%
%
\medskip
\jsection{ADOPTED ACTIVE LEARNING STRATEGY }\par\vspace*{2mm}%
Though exhaustively selecting sample points from all possible grids in the 3D parametric space gives the upmost numerical accuracy, it is nether feasible for large dimension of parameters nor necessary to meet industrial needs. In other words, if ten points are necessary to promise accuracy for each curve fitting, it seems 1,000 simulations are at least required to draw an acceptable conclusion, which is far more time-consuming for research. In contrast, if every simulation is assumed to be independent with one another, which is generally reasonable, and to follow Gaussian Distribution (i.e. the whole set of simulations is assumed to follow a Gaussian Process), which is sound for scenarios without prior information, our study could be simplified to great extent using a supervised algorithm (i.e. Gaussian Process Regression, GPR). In a word, the GP regression in our study would be updated after each simulation, based on which the next sample point with the maximum global uncertainty will be selected.\\

Then we will interpret the above methodology in a mathematical representation, while systematical knowledge in this field are available in many books (Rasmussen et al., 2008; Seeger, 2004). The relationship between the input vector $\mathbf{x}$ and the output vector $\mathbf{y}$ is:
\begin{equation}
  \mathbf{y(x)} = \mathbf{h(x)^T} \mathbf{\beta} + f(x)+\varepsilon    \label{y(x)}
\end{equation}
where, $f(x)\sim GP(0,K(\mathbf{x},\mathbf{x'}))$, $K(\mathbf{x},\mathbf{x'})$ is the kernel function, $\mathbf{h(x)}$ is the standard form of the basis function with coefficients defined as vector $\mathbf{\beta}$, and $\varepsilon \sim N(0,\sigma^2)$ is the measurement error.\\

With this relationship, the marginal likelihood $P(\mathbf{y}|X) = P(\mathbf{y}|\mathbf{f},X)P(\mathbf{f}|X)$ follows a Gaussian distribution $N(H\mathbf{\beta}+\mathbf{f},I)$, i.e. $N(H\mathbf{\beta},K(X,X│\mathbf{\vartheta})+\sigma^2 I)$, where $\mathbf{\vartheta}$ is a hyperparameter vector of the kernel function. As for how to select these unknown latent parameters ($\Vec{\beta}$, $\mathbf{\vartheta}$, $\Vec{\sigma}$) so that the algorithm could find an analytic expression of distribution of the Quantity of Interest (QoI) in the parametric space, a batch algorithm using the negative logarithmic marginal function is usually adopted to spot the most promising candidate (the next sample input $\mathbf{\hat{x}^*}$), which will yield the most information on the whole distribution for next iteration and reduce global maximum uncertainty $\sigma^*(\mathbf{\hat{x}^*})$, while other online adaptive laws using the gradient algorithm or the least-square algorithm are also feasible. Then after the convergence of learning, an explicit distribution function of the GPR-predicted result $\mathbf{y^*}$, with respect to the random input variable $\mathbf{x^*}$ in the parametric space could be formulated, such that the value of QoI at any point in the parametric space is available.
\\

The selection of the kernel function $K(\mathbf{x},\mathbf{x'})$ and the basis function $\vec{h(x)}$ of GPR holds the key to learning and convergence speed in many cases. How to determine the most appropriate kernel function as well as related hyperparameters is still a heated field with possible algorithms including Deep Belief Net (Salakhutdinov et al., 2008), Structure Search Method (Duvenaud et al., 2008) and Hierarchical Bayesian Framework (Schwaighofer et al., 2005), while these methods are all similar in essence that is to minimize the uncertainty. Based on the hydrodynamic properties of our target coefficients and referring existing studies on kernel selection in fluid mechanics (Fan et al., 2019), Pure Quadratic function was chosen as our basis function and ARD Matern 3/2 was picked as our kernel function.\\

\begin{figure}[htb]
\centering  
    \begin{subfigure}[b]{\linewidth}
        \centering
       \includegraphics[width=\linewidth]{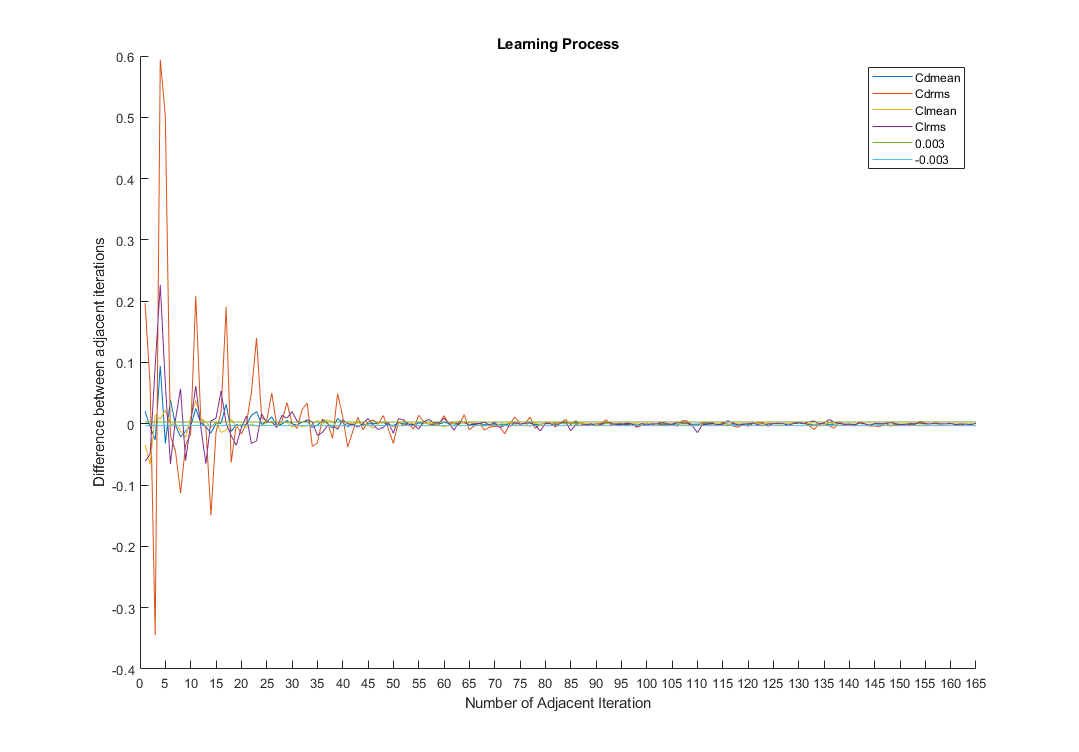}%
        \caption{Adjacent difference from the first to the last}
    \end{subfigure}
    \vskip\baselineskip
    \begin{subfigure}[b]{\linewidth}
        \centering
        \includegraphics[width=\linewidth]{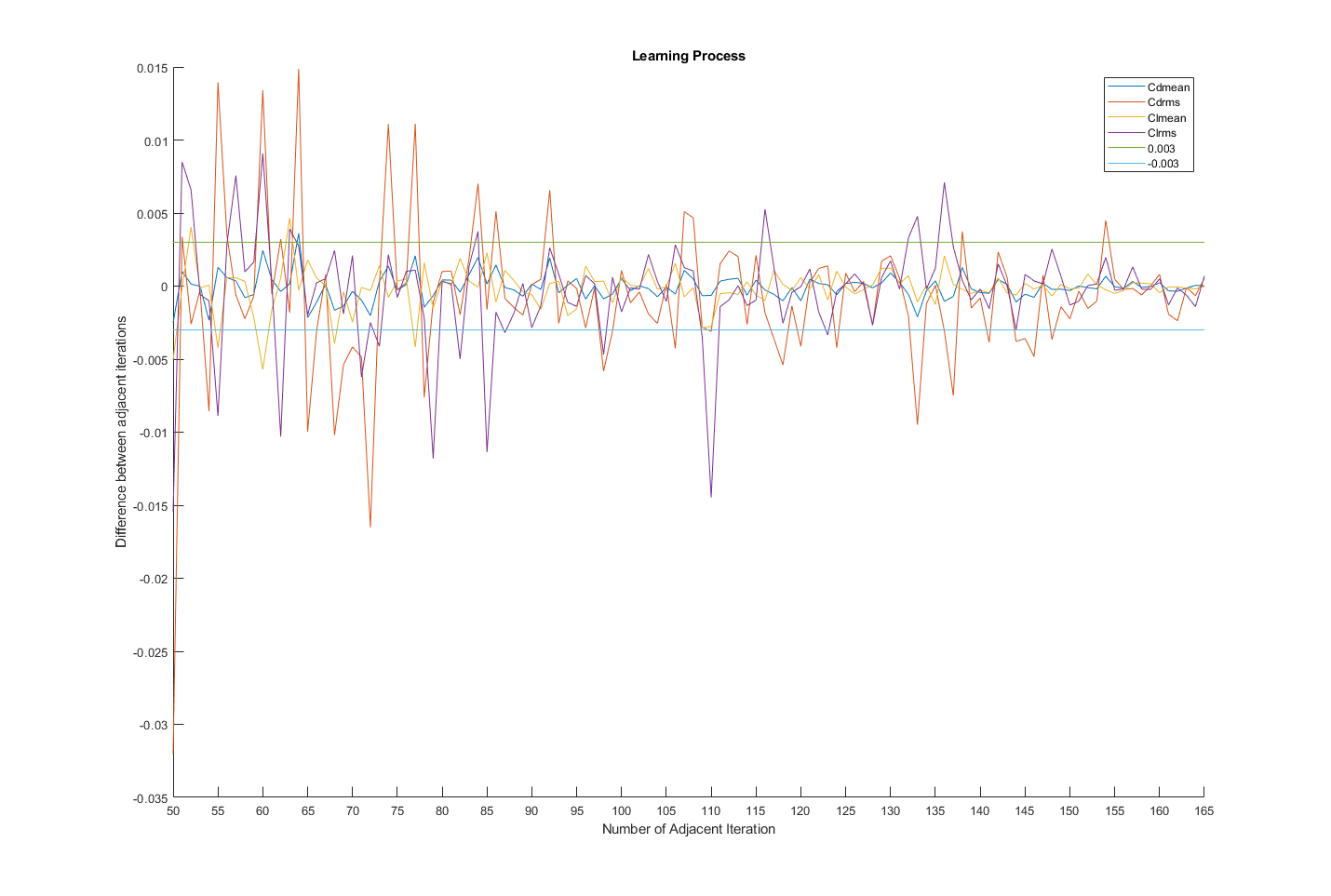}
        \caption{Adjacent difference from No. 50 - 51 to the last}
    \end{subfigure}
\caption{Convergence process}
\label{Fig:diff}
\end{figure}

\begin{figure}[htb]
\centering
\includegraphics[width=\linewidth]{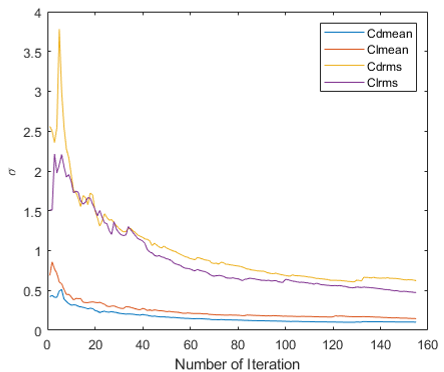}
\caption{Standard deviations} \label{Fig:std}
\end{figure}

\begin{figure}[htb]
\centering  
    \begin{subfigure}[b]{\linewidth}
        \centering
       \includegraphics[width=\linewidth,height=0.2\linewidth]{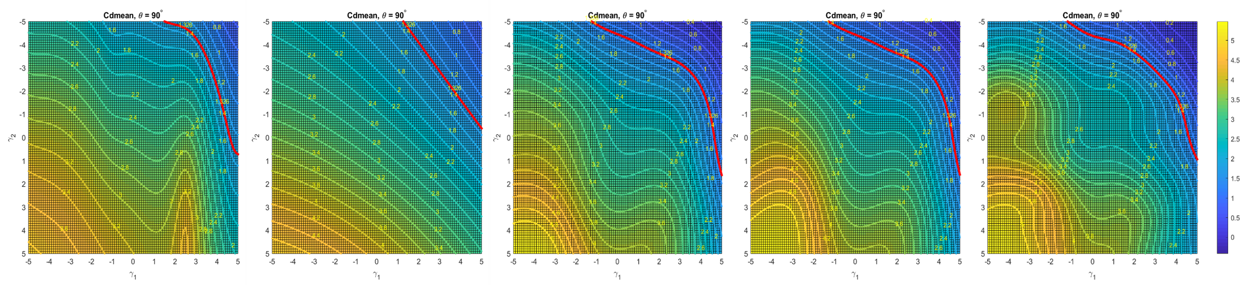}%
        \caption{$C_D^{MEAN}$: iteration 4, 5, 23, 24, last}
    \end{subfigure}
    \vskip\baselineskip
    \begin{subfigure}[b]{\linewidth}
        \centering
        \includegraphics[width=\linewidth,height=0.2\linewidth]{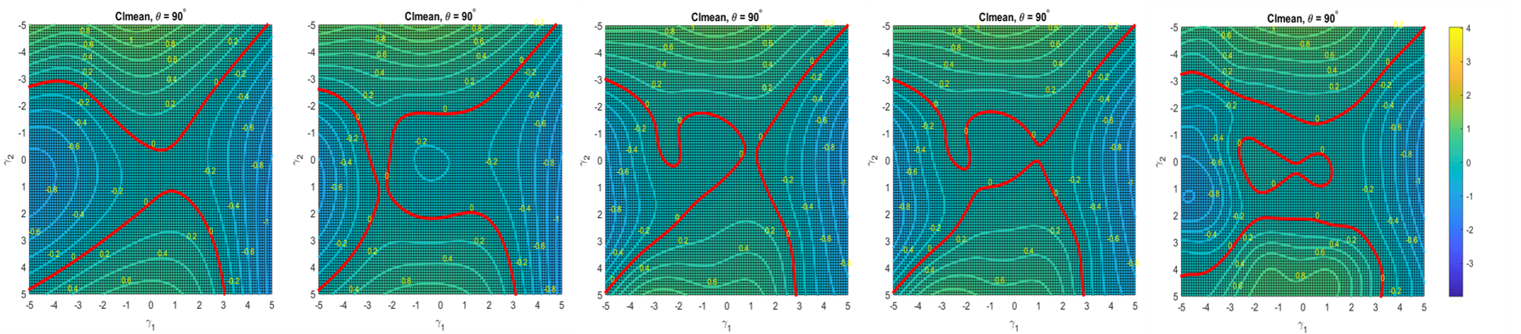}
        \caption{$C_L^{MEAN}$: iteration 11, 12, 63, 64, last}
    \end{subfigure}
    \vskip\baselineskip
    \begin{subfigure}[b]{\linewidth}
        \centering
        \includegraphics[width=\linewidth,height=0.2\linewidth]{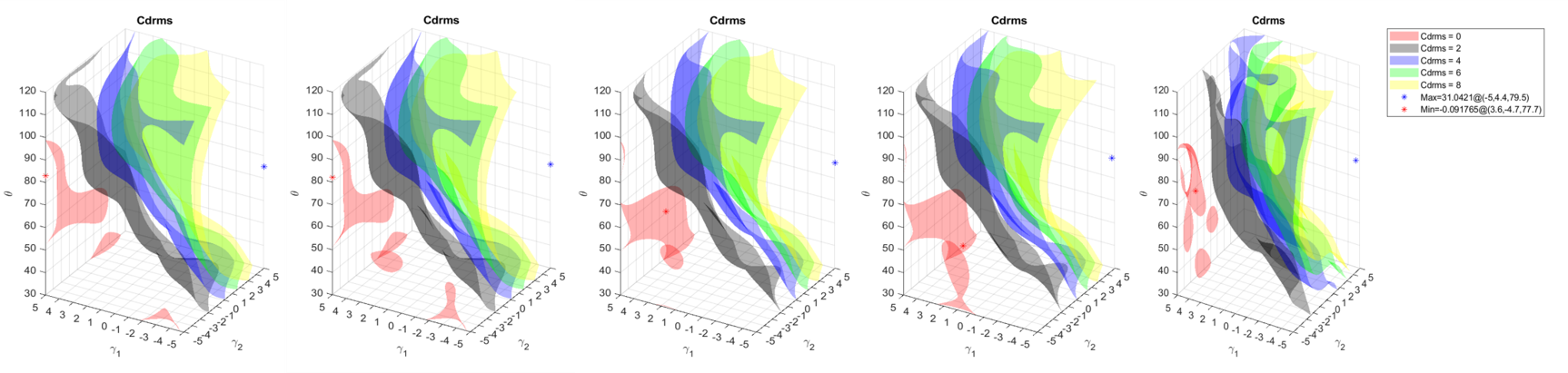}
        \caption{$C_D^{RMS}$: iteration 11, 12, 16, 17, last}
    \end{subfigure}
    \vskip\baselineskip
    \begin{subfigure}[b]{\linewidth}
        \centering
        \includegraphics[width=\linewidth,height=0.2\linewidth]{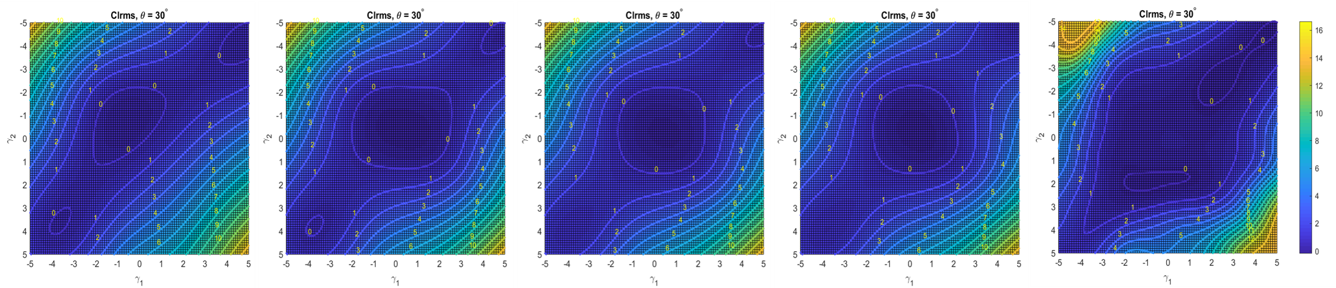}
        \caption{$C_L^{RMS}$: iteration 13, 14, 16, 17, last}
    \end{subfigure}
\caption{Remarkable jumps in learning process}
\label{Fig:learning}
\end{figure}

For our four hydrodynamic coefficients ($C_D^{MEAN}$, $C_L^{MEAN}$, $C_D^{RMS}$, $C_L^{RMS}$), four Gaussian Process Regressions were conducted simultaneously. Numerical simulations started from twenty randomly selected sample points in the three-parameter space and continued with sequential sample points recommended by GPR. Combined with the calculation of four mean values for four coefficients of 2000 randomly predetermined points in the parametric space at each iteration, the convergence criterion in our study was defined as successive 10 norms of the difference between two sequential mean values are all bounded by 0.003, a relatively small number, for each coefficient. The convergence of learning means the information gathering procedure becomes almost saturate, i.e. the machine has basically mastered the physics underneath simulation results so that it is time to terminate iterations. The learning process in our study is presented in figure \ref{Fig:diff} and the learning process of four QoI distributions are demonstrated as figure \ref{Fig:learning}, from which it’s obvious that where there is a jump, there is a remarkable learning breakthrough in the progress of GPR. In addition, the decrease of four standard deviations in our study are demonstrated as figure \ref{Fig:std}, implying the learning process is by no means linear and monotonic but an overall downward and asymptotic tendency provided kernel function is selected appropriately.\\


%
%
\jsection{EFFECTS OF PARAMETERS}\par\vspace*{2mm}%

\begin{figure}[htb]
\centering
\includegraphics[width=\linewidth]{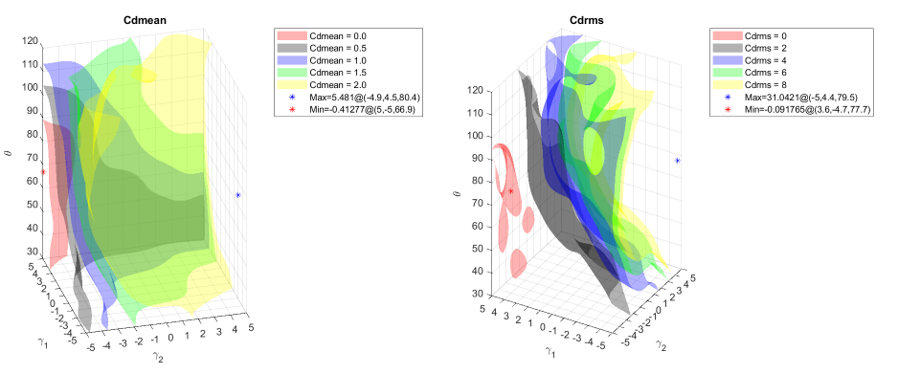}
\includegraphics[width=\linewidth]{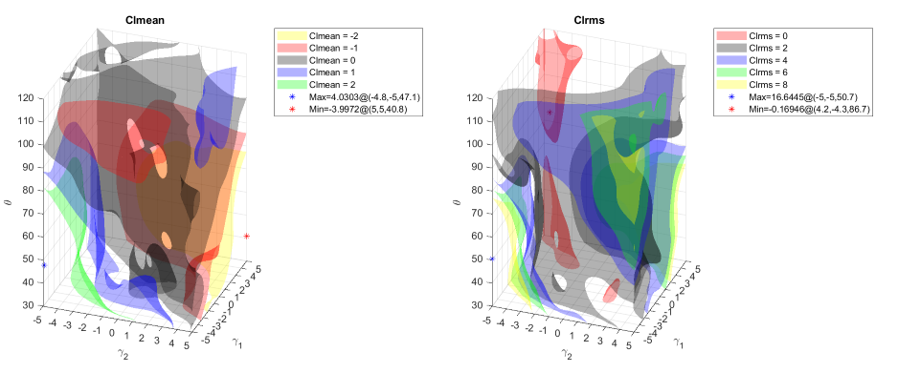}
\caption{Global distributions of $C_D^{MEAN}$, $C_L^{MEAN}$, $C_D^{RMS}$, $C_L^{RMS}$} \label{Fig:global}
\end{figure}

After the convergence of all four GP regressions, we plotted the global distributions of four hydrodynamic coefficients in the 3D-parametric space (Figure \ref{Fig:global}), from which diverse flow control strategies for different control goals could be designed, and the specific effects of these three parameters will be discussed in the following part.\\

From the global distribution of $C_D^{MEAN}$, though its maximum value shows up at a relatively high angle, the minimum one is rather far from the separation point with a negative value, indicating the finished transform from drag to thrust with this promising model. Apart from the extreme values, the steep isosurface of $C_D^{MEAN}$ = 0 implies less sensitivity of QoI to the relative angle $\theta$ than that to rotation rates $\gamma_{1,2}$. As for $C_L^{MEAN}$, in significant contrast to the continuous isosurfaces of $C_D^{MEAN}$, it has disjoint regions, especially at the widely extended isosurface of $C_L^{MEAN}$ = 0, which has two symmetric planes perpendicular to each other. As for the global distributions of RMS values, they show similar independence on the relative angle after control cylinders placed higher than 90$^{\circ}$, though $\theta$ still plays a significant role among lower angles. However, the maximum and the minimum values of $C_L^{RMS}$ both locate at low angles, remarkably distinguished from those of $C_D^{RMS}$, also with its significantly lower maximum value than $C_L^{RMS}$ indicating there are more drastic vibrations along the coming flow direction than those along the perpendicular direction.
\\
    
\begin{figure}[htb]
\centering
\includegraphics[width=\linewidth]{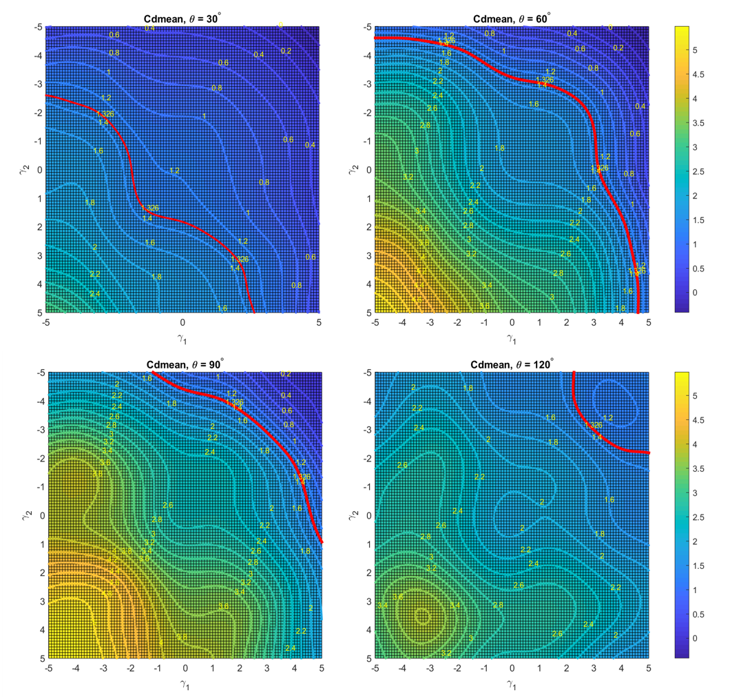}
\caption{Effects of $\gamma_1$ and $\gamma_2$ on $C_D^{MEAN}$ with $\theta$ = 30$^{\circ}$,60$^{\circ}$,90$^{\circ}$,120$^{\circ}$} 
\label{Fig:effcdmeanR}
\end{figure}

\begin{figure}[htb]
\centering
\includegraphics[width=\linewidth]{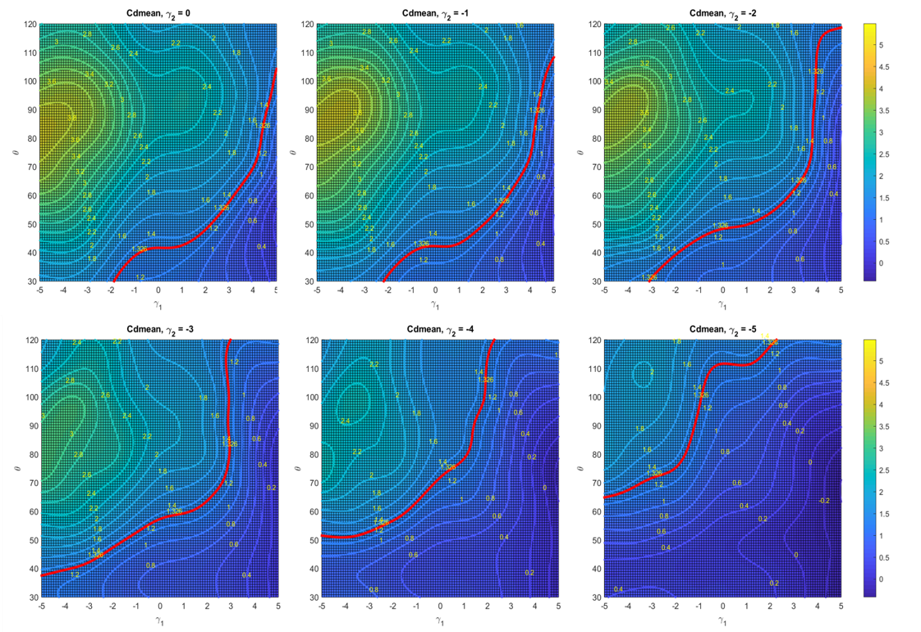}
\caption{Effects of $\theta$ and $\gamma_1$ on $C_D^{MEAN}$ with $\gamma_2$ = 0,-1,-2,-3,-4,-5} 
\label{Fig:effcdmeanTH}
\end{figure}

From figure \ref{Fig:effcdmeanR}, the lowest value of $C_D^{MEAN}$ comes up with two cylinders counter-rotating inwardly at the maximum rate, while the highest one happens when they rotating at the maximum rate outwardly. Besides, this $C_D^{MEAN}$ distribution is basically symmetrical along the diagonal, indicating the effects of two rotation rates ($\gamma_1$, $\gamma_2$) are similar, but what should be noticed here is the value of both axes increases with respect to a faster inward rotation rate, vice versa. The reason why their effects are not strictly symmetrical mainly stems from the asymmetry of the kernel function, though this influence is insignificant to affect the final conclusion. Furthermore, the growing effectiveness of two rotation rates also manifests itself in the increasingly denser contours at 30$^{\circ}$ to 90$^{\circ}$ in figure \ref{Fig:effcdmeanR}, whereas the control effective region (in our study defined as the region within the red line at 1.328, which is the $C_D^{MEAN}$ of isolated cylinder without flow control) shrinks significantly. Similarly, the effective area converts to the effective width when only one $\gamma$ focused, like figure \ref{Fig:effcdmeanTH}, where this width decreases before the angle reaching roughly 80$^{\circ}$, and remains basically unchanged till control cylinders too far (like 120$^{\circ}$) from the separation point to make a difference to the wake of the main cylinder unless their effects are enhanced enough by injecting energy and speeding up rates (at least around 2.5 from the last graph in figure \ref{Fig:effcdmeanR}). What also could be concluded from the control effective zone characterised by crossing all $\gamma_2$ (bottom three in figure \ref{Fig:effcdmeanTH}) is that one cylinder becomes so powerful when rotating faster than a threshold that even making its counterpart trivial. This threshold readable from figure \ref{Fig:effcdmeanR} tend to grow with increasing relative angles: 2.5 at 30$^{\circ}$, 4.6 at 60$^{\circ}$ and greater than 5 at 90$^{\circ}$. This rising trend not only demonstrates that the drag reduction is always achievable via this model as long as the arrangement is suitable, but also makes its working width broad enough to benefit practical applications where ubiquitous unexpected scenarios may result in the breakdown of one actuator. \\

The coupled effect between $\theta$ and $\gamma$ also shows up in figure \ref{Fig:effcdmeanTH}, where the relative angle for the lowest $C_D^{MEAN}$ increases from 45$^{\circ}$ to 70$^{\circ}$ with faster control cylinder rates. However, without energy input ($\gamma_1$ = $\gamma_2$ = 0) these two small cylinders are only effective when located lower than 40$^{\circ}$ (Figure \ref{Fig:effcdmeanTH}(a)), after which they will militate against the task.
\\

\begin{figure}[htb]
\centering
\includegraphics[width=\linewidth]{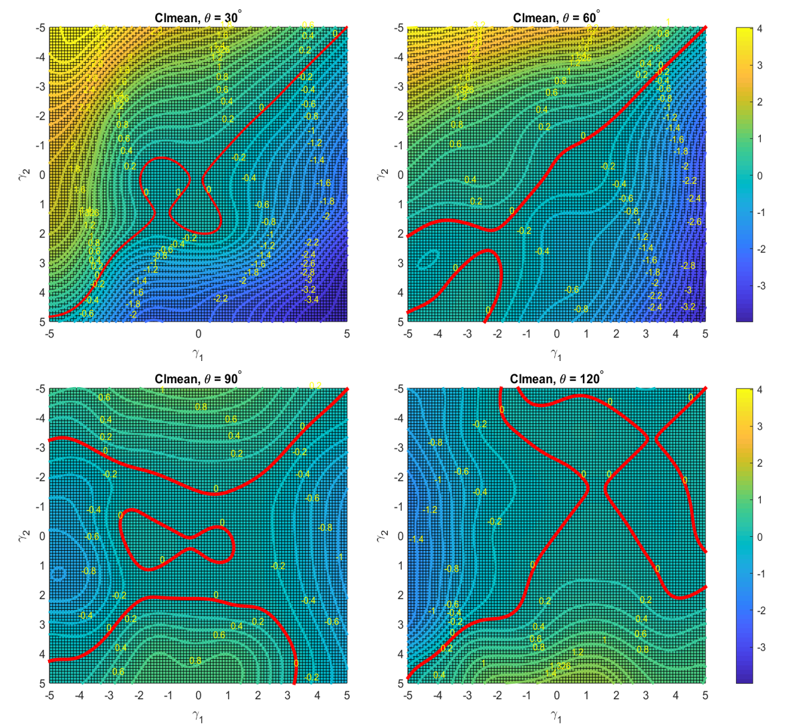}
\caption{Effects of $\gamma_1$ and $\gamma_2$ on $C_L^{MEAN}$ with $\theta$ = 30$^{\circ}$,60$^{\circ}$,90$^{\circ}$,120$^{\circ}$} 
\label{Fig:effclmeanR}
\end{figure}

In significant contrast to $C_D^{MEAN}$, $C_L^{MEAN}$ stays roughly zero as two cylinders counter-rotating, and reaches relative high values at a wide range of $\theta$. Moreover, though two actuators seem to have higher effectiveness at lower angles (bright, dense and steep contour lines in figure \ref{Fig:effclmeanR}(a),(b)) with the brightest color appearing when the rotation directions same, they turn out to be almost useless in 90$^{\circ}$ and 120$^{\circ}$ (roughly monochrome contour maps as figure \ref{Fig:effclmeanR}(c),(d)) and somehow demonstrate symmetry along the diagonal line in figure \ref{Fig:effclmeanR}(c). 
\\

\begin{figure}[htb]
\centering
\includegraphics[width=\linewidth]{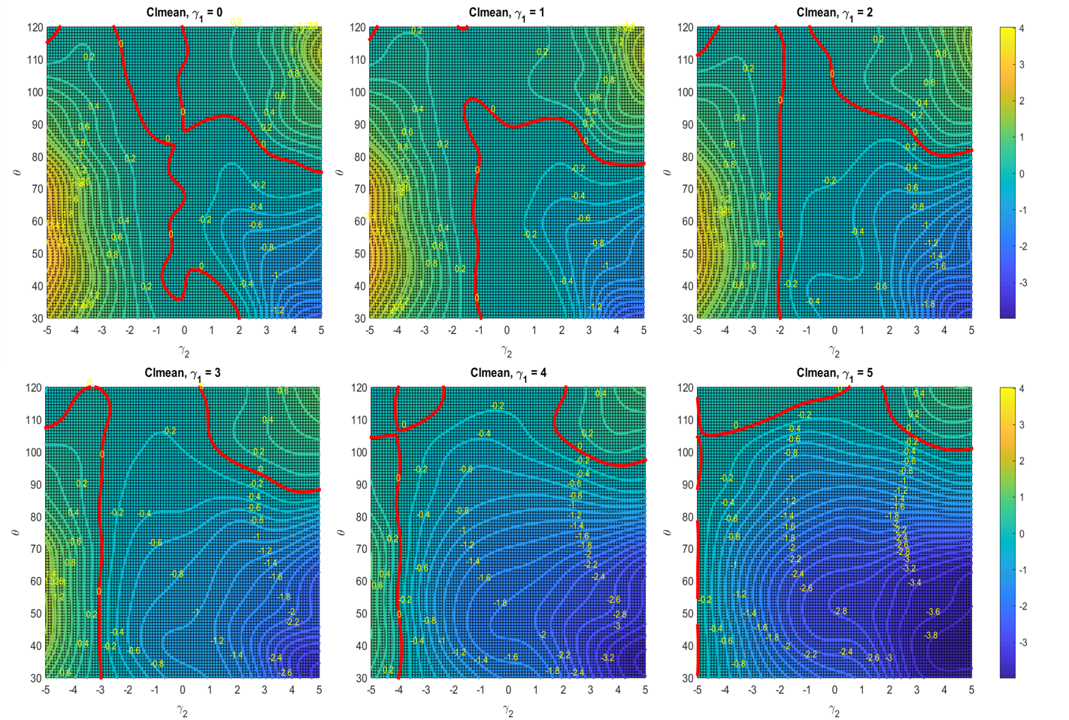}
\caption{Effects of $\theta$ and $\gamma_2$ on $C_L^{MEAN}$ with $\gamma_1$ = 0,1,2,3,4,5} 
\label{Fig:effclmeanTH}
\end{figure}

To differentiate effects of the relative angle and rotation rates, predetermining one rotation rate as $\gamma_i$ (i = 1 or 2), the control effect seems to be solely dependent on the relative angle when $\theta$ is around 80$^{\circ}$ and the other cylinder rotating along identical direction at a rate $\gamma_j$ (j = 2 or 1, $j \neq i$) higher than 3 (Figure \ref{Fig:effclmeanTH}). This overwhelming domination of one parameter, where the contours are extraordinary dense and almost perpendicular to that parametric axis is defined as the dominant zone of that parameter in our study, indicating the other parameter corresponding to the other axis loses its effects completely. In contrast, for $\theta$ less than 80$^{\circ}$ and $\gamma_j$ greater than 3 with opposite sign of $\gamma_i$, the dominant zone turns to belong to $\gamma_j$. The above discovery paves the way for choosing the most efficient strategy to achieve control goals better by modifying the value of appropriate parameter, at best to its dominant range, at worst avoiding to that of its counterpart. \\

\begin{figure}[htb]
\centering
\includegraphics[width=\linewidth]{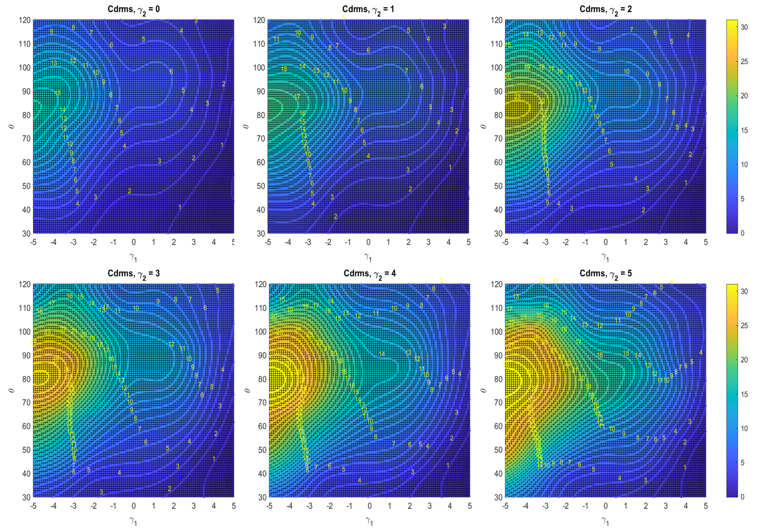}
\caption{Effects of $\theta$ and $\gamma_1$ on $C_D^{RMS}$ with $\gamma_2$ = 0,1,2,3,4,5} 
\label{Fig:effcdrmsTH}
\end{figure}

\begin{figure}[htb]
\centering
\includegraphics[width=\linewidth]{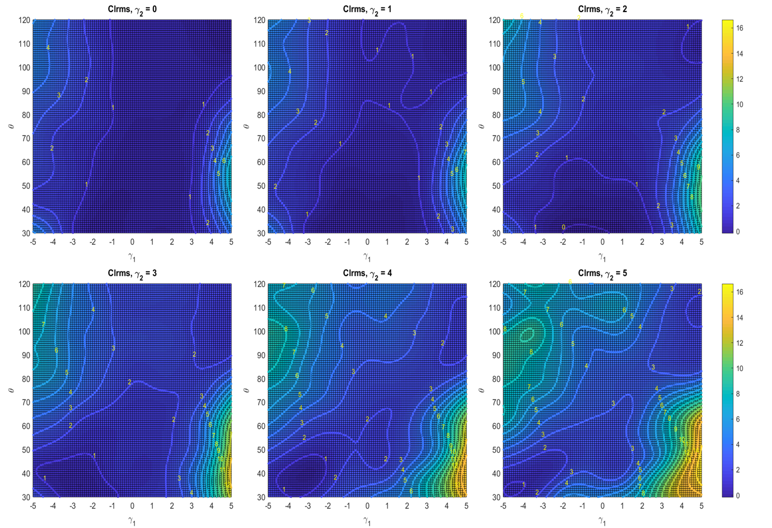}
\caption{Effects of $\theta$ and $\gamma_1$ on $C_L^{RMS}$ with $\gamma_2$ = 0,1,2,3,4,5} 
\label{Fig:effclrmsTH}
\end{figure}

As for $C_D^{RMS}$, although similar with $C_D^{MEAN}$ its smaller values come up with two control cylinders rotating inwardly (Figure \ref{Fig:global}), it highly depends on $\theta$ (Figure \ref{Fig:effcdrmsTH}), not only as the maximum value always appears around 80$^{\circ}$, but almost the whole achievable range of $\theta$ becomes dominant zone after outward rotation rates of two cylinders greater than 3. However, different from $C_D^{RMS}$, $C_L^{RMS}$ is less influenced by our three parameters, showing obviously narrower warm-color regions and lower maximum values than those of $C_D^{RMS}$ (Figure \ref{Fig:effclrmsTH}), though they have similar $\theta$-dependent characteristic when $\theta$ is greater than 80$^{\circ}$ and two rotation rates are small. From the above discovery, the necessity of collaborations between two control cylinders is a distinction of $C_L^{RMS}$: with one rotation rate lower than 3, this model is useless no matter how fast the other cylinder rotates towards either side, which highlights the importance of taking care of both actuators to maintain desirable effects for $C_L^{RMS}$-targeted control system. 
\\
%
%
%
\jsection{mechanism}\par\vspace*{2mm}%
In this section, we will analyze the flow patterns of our model and discuss the fluid mechanism underneath the hydrodynamic characteristics described in previous sections.\\

\begin{figure}[htb]
\centering  
    \begin{subfigure}[b]{0.5\linewidth}
        \includegraphics[width=\linewidth]{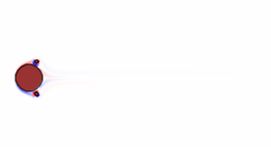}
        \caption{$\gamma_1$ = 5, $\gamma_2$ = -5, $\theta$ = 65$^{\circ}$}
    \end{subfigure}%
    \begin{subfigure}[b]{0.5\linewidth}
        \includegraphics[width=\linewidth]{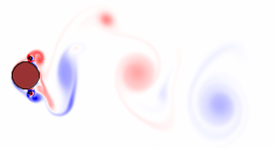}
        \subcaption{$\gamma_1$ = -5, $\gamma_2$ = 5, $\theta$ = 75$^{\circ}$}
    \end{subfigure}%
\caption{Extreme values of $C_D^{MEAN}$\\
(a) $C_D^{MEAN}$= -0.42, $C_D^{RMS}$= 0.17, $C_L^{MEAN}$= 0, $C_L^{RMS}$= 0\\
(b) $C_D^{MEAN}$= 5.41,$C_D^{RMS}$= 30.53, $C_L^{MEAN}$= 0.15, $C_L^{RMS}$= 7.93}
\label{Fig:cdmean}
\end{figure}

\begin{figure}[htb]
\centering  
    \begin{subfigure}[b]{0.5\linewidth}
        \includegraphics[width=\linewidth]{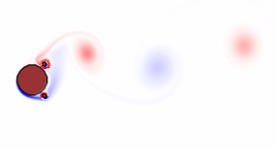}
        \caption{$\gamma_1$ = -5, $\gamma_2$ = -5, $\theta$ = 52.5$^{\circ}$}
    \end{subfigure}%
    \begin{subfigure}[b]{0.5\linewidth}
        \includegraphics[width=\linewidth]{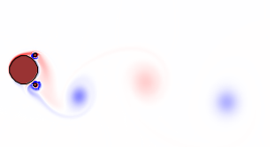}
        \caption{$\gamma_1$ = 5, $\gamma_2$ = 5, $\theta$ = 52.5$^{\circ}$}
    \end{subfigure} \\
    \begin{subfigure}[b]{0.5\linewidth}
        \includegraphics[width=\linewidth]{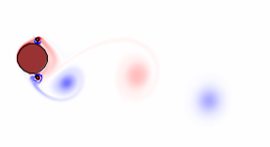}
        \caption{$\gamma_1$ = 5, $\gamma_2$ = 5, $\theta$ = 75$^{\circ}$}
    \end{subfigure}%
    \begin{subfigure}[b]{0.5\linewidth}
        \includegraphics[width=\linewidth]{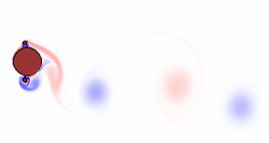}
        \caption{$\gamma_1$ = 5, $\gamma_2$ = 5, $\theta$ = 97.5$^{\circ}$}
    \end{subfigure}%
\caption{Asymmetric wake zone\\
(a) $C_D^{MEAN}$= 0.77, $C_D^{RMS}$= 0.67, $C_L^{MEAN}$= 3.92, $C_L^{RMS}$= 15.83\\
(b) $C_D^{MEAN}$= 0.76,$C_D^{RMS}$= 0.65,$C_L^{MEAN}$= -3.87,$C_L^{RMS}$= 15.48\\
(c) $C_D^{MEAN}$= 1.50, $C_D^{RMS}$= 2.53, $C_L^{MEAN}$= -2.77, $C_L^{RMS}$= 8.59\\
(d) $C_D^{MEAN}$= 1.81, $C_D^{RMS}$= 3.45, $C_L^{MEAN}$= -0.10, $C_L^{RMS}$= 2.22}
\label{Fig:cl}
\end{figure}

From the comparison of high and low values of $C_D^{MEAN}$ (Figure \ref{Fig:cdmean}), it could be verified that the drag reduction for bluff body results from the flow reattachment after flow separation point, and the viscous effect of two inward fast-rotating cylinders at low relative angles makes it more effective to stabilize the wake of main cylinder. In contrast, high values of $C_D^{MEAN}$ represent more flow disturbance due to more vortexes aroused by outward rotations of two actuators at relative angle around the separation point. Different from the maximum value of $C_D^{MEAN}$ always coming along with counter-rotating cylinders, the highest vibration amplitude and intensity in lift direction both show up with actuators rotating along the same direction (Figure \ref{Fig:cl}(a)), which irritates the wake and lead to a larger perpendicular component of hydrodynamic force.\\

In order to explain the completely $\theta$-dependent characteristic of lift coefficients when the relative angle greater than 80$^{\circ}$, we analyzed flow patterns in figure \ref{Fig:cl}(b)-(d), where it is around 80$^{\circ}$ that the positive vorticity from the upper cylinder covers the buttock of the main cylinder exactly. This angle is obviously independent of two rotation rates but may relate to the predetermined parameters such as the normalised gap and the diameter ratio. Though there is no analytic solution to find this crucial angle, for those who encounter problems to determine the control effective zone of $\theta$ for lift coefficients we provide a reference on how to find this critical angle as purposely checking the flow patterns. Furthermore, for those who endeavor to find a cost-efficient way achieving drag reduction there is also a reference (Figure \ref{Fig:cl}(d)), where one cylinder with suitable rotation rates is already sufficient to reduce $C_D^{MEAN}$ at small relative angles even with the worst partner.\\

\begin{figure}[htb]
\centering  
\includegraphics[width=\linewidth]{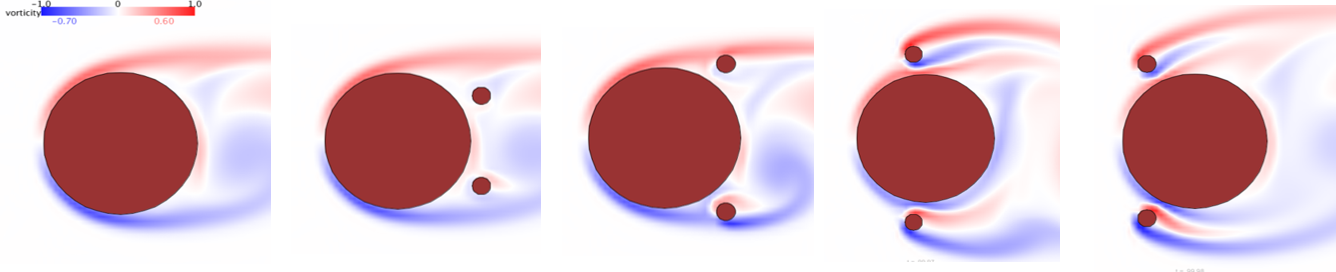}
\caption{Relative location with respect to separation point\\
from left to right: ($\gamma = 0$) \\
isolate cylinder, $\theta$ = 30$^{\circ}$, 52.5$^{\circ}$, 97.5$^{\circ}$, 120$^{\circ}$\\
$C_D^{MEAN}$ = 1.378, 1.21, 1.57, 2.48, 1.93
}
\label{Fig:sepa}
\end{figure}

\begin{figure}[htb]
\centering  
    \begin{subfigure}[b]{0.5\linewidth}
        \includegraphics[width=\linewidth]{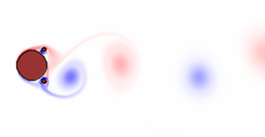}
        \caption{$\gamma_1$ = 2.5, $\gamma_2$ = 5, $\theta$ = 52.5$^{\circ}$}
    \end{subfigure}%
    \begin{subfigure}[b]{0.5\linewidth}
        \includegraphics[width=\linewidth]{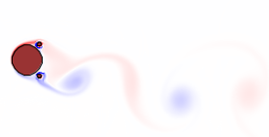}
        \caption{$\gamma_1$ = 5, $\gamma_2$ = 2.5, $\theta$ = 52.5$^{\circ}$}
    \end{subfigure}%
\caption{Effect of certain side cylinder on $C_L^{RMS}$\\
(a) $C_D^{MEAN}$= 2.32, $C_D^{RMS}$= 5.59, $C_L^{MEAN}$= -1.93, $C_L^{RMS}$= 4.93\\
(b) $C_D^{MEAN}$= 0.45,$C_D^{RMS}$= 0.21, $C_L^{MEAN}$=-3.54, $C_L^{RMS}$=12.59}
\label{Fig:clrms}
\end{figure}

Besides, the reason why the closer two actuators arranged towards the separation point, the more sensitive two drag coefficients become with respect to rotation rates attributes to the growing effectiveness of momentum injection from control cylinders (Figure \ref{Fig:sepa}). On the other hand, the reason why $C_L^{RMS}$ increases remarkably with merely one certain control cylinder rotating faster than a threshold rate (around 3 from contour maps with limited warm color zone in figure \ref{Fig:effclrmsTH}) lies in the vorticity generated from control cylinders (Figure \ref{Fig:clrms} and Figure \ref{Fig:cl}(b)). Therefore, instead of improving the rate of cylinder rotating outwardly, it is more sensible to speed up the rate of the other rotating inwardly so that there are more momentum crossing the wake of main cylinder, where vortex shedding derives a huge amount of momentum from vortexes generated by actuators and leads to insignificant lift force. \\

%
\jsection{conclusion}\par\vspace*{2mm}%
In this paper, the flow control problem of the flow passing a stationary cylinder at a fixed Reynold number 500 using two control cylinders with different rotation rates was analyzed via an active learning strategy based on GPR. Regression results of four hydrodynamic coefficients, the mean and the RMS values of lift and drag force, were systematically analyzed with respect to three parameters, relative angle of control cylinders to the main cylinder and two different rotation rates. Based on numerical simulations, the mechanism underneath the parametric effects including the control effective zone and the dominant zone of different parameters was further studied by analyzing flow patterns, such that more effective and efficient methods to improve the hydrodynamic characteristics of the target body were promoted. 

\jsection{REFERENCES}\par\vspace*{2mm}%

\begin{minipage}[t]{86.5mm}
\noindent\hspace*{-5.2mm} %
Cheng, M and Luo, LS (2007). %
\lq\lq Characteristics of two-dimensional flow around a rotating circular cylinder near a plane wall\rq\rq, {\it Physics of Fluids},
Vol.\,19, No\,\,6, pp.\,063601.
\end{minipage}\hfill
\\\vspace*{1.5mm}\\
\begin{minipage}[t]{86.5mm}
\noindent\hspace*{-5.2mm} %
Choi, H, Jeon, WP and Kim, J (2008). %
\lq\lq Control of flow over a bluff body\rq\rq, {\it Annu. Rev. Fluid Mech.},
Vol.\,40, pp.\,113--139.
\end{minipage}\hfill
\\\vspace*{1.5mm}\\
\begin{minipage}[t]{86.5mm}
\noindent\hspace*{-5.2mm} %
Duvenaud, D,  Lloyd, JR, Grosse, R, Tenenbaum, JB and Ghahramani, Z (2013). %
\lq\lq Structure discovery in nonparametric regression through compositional kernel search\rq\rq, {\it arXiv preprint arXiv:1302.4922}.
\end{minipage}\hfill
\\\vspace*{1.5mm}\\
\begin{minipage}[t]{86.5mm}
\noindent\hspace*{-5.2mm} %
Fan, D, Jodin, G, Consi, TR, Bonfiglio, L, Ma, Y, Keyes, LR, Karniadakis, GE and Triantafyllou, MS (2019). %
\lq\lq A robotic Intelligent Towing Tank for learning complex fluid-structure dynamics\rq\rq, {\it Science Robotics},
Vol.\,4, No\,\,36.
\end{minipage}\hfill
\\\vspace*{1.5mm}\\
\begin{minipage}[t]{86.5mm}
\noindent\hspace*{-5.2mm}
Hao, M, Thapa, J, Cheng, L, Zhou, T. (2013). %
\lq\lq Three-dimensional transition of vortex shedding flow around a circular cylinder at right and oblique attacks\rq\rq, {\it Physics of Fluids}, Vol.\,25, No\,\,1, pp.\,014105.
\end{minipage}\hfill
\\\vspace*{1.5mm}\\
\begin{minipage}[t]{86.5mm}
\noindent\hspace*{-5.2mm}
Henderson, RD (1995). %
\lq\lq Details of the drag curve near the onset of vortex shedding\rq\rq, {\it Physics of Fluids}, Vol.\,7, No\,\,9, pp.\,2102--2104.
\end{minipage}\hfill
\\\vspace*{1.5mm}\\
\begin{minipage}[t]{86.5mm}
\noindent\hspace*{-5.2mm} %
Hinton, GE and Salakhutdinov, RR (2008). %
\lq\lq Using deep belief nets to learn covariance kernels for Gaussian processes\rq\rq, {\it Advances in neural information processing systems}, pp.\,1249--1256.
\end{minipage}\hfill
\\\vspace*{1.5mm}\\
\begin{minipage}[t]{86.5mm}
\noindent\hspace*{-5.2mm}
Jahanmiri, M (2010). %
\lq\lq Active flow control: a review \rq\rq, {\it
Chalmers University of Technology}
\end{minipage}\hfill
\\\vspace*{1.5mm}\\
\begin{minipage}[t]{86.5mm}
\noindent\hspace*{-5.2mm}
Ji, C, Zhang, Z, Xu, D, Srinil, N. (2019). %
\lq\lq Three-Dimensional Direct Numerical Simulations of Flows Past an Inclined Cylinder Near a Plane Boundary\rq\rq, {\it ASME 2019 38th International Conference on Ocean, Offshore and Arctic Engineering}.
\end{minipage}\hfill
\\\vspace*{1.5mm}\\
\begin{minipage}[t]{86.5mm}
\noindent\hspace*{-5.2mm}
Jiang, HY, Cheng, L (2017). %
\lq\lq Strouhal--Reynolds number relationship for flow past a circular cylinder\rq\rq, {\it Journal of Fluid Mechanics}, Vol.\,832, pp.\,170--188.
\end{minipage}\hfill
\\\vspace*{1.5mm}\\
\begin{minipage}[t]{86.5mm}
\noindent\hspace*{-5.2mm}
Kang, S (2003). %
\lq\lq Characteristics of flow over two circular cylinders in a side-by-side arrangement at low Reynolds numbers\rq\rq, {\it Physics of Fluids}, Vol.\,15, No\,\,9, pp.\,2486--2498.
\end{minipage}\hfill
\\\vspace*{1.5mm}\\
\begin{minipage}[t]{86.5mm}
\noindent\hspace*{-5.2mm}
Kumar, RA, Sohn, CH, Gowda, BH. (2008). %
\lq\lq Passive control of vortex-induced vibrations: an overview\rq\rq, {\it Recent Patents on Mechanical Engineering}, Vol.\,1, No\,\,1, pp.\,1--11.
\end{minipage}\hfill
\\\vspace*{1.5mm}\\
\begin{minipage}[t]{86.5mm}
\noindent\hspace*{-5.2mm} %
Kuo, CH, Chiou, LC, Chen, CC (2007)%
\lq\lq Wake flow pattern modified by small control cylinders at low Reynolds number \rq\rq, {\it Journal of Fluids and Structures},
Vol.\,23, No\,\,6, pp.\,938--956.
\end{minipage}\hfill
\\\vspace*{1.5mm}\\
\begin{minipage}[t]{86.5mm}
\noindent\hspace*{-5.2mm} %
Mittal, S (2001). %
\lq\lq Control of flow past bluff bodies using rotating control cylinders \rq\rq, {\it Journal of Fluids and Structures},
Vol.\,15, No\,\,2, pp.\,291--326.
\end{minipage}\hfill
\\\vspace*{1.5mm}\\
\begin{minipage}[t]{86.5mm}
\noindent\hspace*{-5.2mm} %
Mittal, S (2003). %
\lq\lq Flow control using rotating cylinders: effect of gap \rq\rq, {\it J. Appl. Mech.},
Vol.\,70, No\,\,5, pp.\,762--770.
\end{minipage}\hfill
\\\vspace*{1.5mm}\\
\begin{minipage}[t]{86.5mm}
\noindent\hspace*{-5.2mm} %
Modi, VJ (1997). %
\lq\lq Moving surface boundary-layer control: A review \rq\rq, {\it Journal of Fluids and Structures},
Vol.\,11, No\,\,6, pp.\,627--663.
\end{minipage}\hfill
\\\vspace*{1.5mm}\\
\begin{minipage}[t]{86.5mm}
\noindent\hspace*{-5.2mm}
Muddada, S, Patnaik, BSV (2010). %
\lq\lq An active flow control strategy for the suppression of vortex structures behind a circular cylinder\rq\rq, {\it Physics of Fluids}, Vol.\,29, No\,\,2, pp.\,93--104.
\end{minipage}\hfill
\\\vspace*{1.5mm}\\
\begin{minipage}[t]{86.5mm}
\noindent\hspace*{-5.2mm} %
Past{\`o}, S (2008). %
\lq\lq Vortex-induced vibrations of a circular cylinder in laminar and turbulent flows\rq\rq, {\it Journal of Fluids and Structures},
Vol.\,24, No\,\,7, pp.\,977--993.
\end{minipage}\hfill
\\\vspace*{1.5mm}\\
\begin{minipage}[t]{86.5mm}
\noindent\hspace*{-5.2mm} %
Rao, A, Stewart, BE, Thompson, MC, Leweke, T, Hourigan, K. (2011). %
\lq\lq Flows past rotating cylinders next to a wall\rq\rq, {\it Journal of Fluids and Structures},
Vol.\,27, No\,\,5-6, pp.\,668--679.
\end{minipage}\hfill
\\\vspace*{1.5mm}\\
\begin{minipage}[t]{86.5mm}
\noindent\hspace*{-5.2mm} %
Rasmussen, CE and Williams, CK (2006). %
\lq\lq Gaussian processes for machine learning, vol. 1\rq\rq, {\it MIT press},
Vol.\,39, pp.\,40--43.
\end{minipage}\hfill
\\\vspace*{1.5mm}\\
\begin{minipage}[t]{86.5mm}
\noindent\hspace*{-5.2mm} %
Sakamoto, H, Haniu, H. (1994). %
\lq\lq Optimum Suppression of Fluid Forces Acting on a Circular Cylinder \rq\rq, {\it Journal of Fluids Engineering},
Vol.\,116, No\,\,2, pp.\,221–227.
\end{minipage}\hfill
\\\vspace*{1.5mm}\\
\begin{minipage}[t]{86.5mm}
\noindent\hspace*{-5.2mm} %
Sarpkaya, T (1979). %
\lq\lq Vortex-induced oscillations: a selective review\rq\rq, {\it Journal of applied mechanics},
Vol.\,46, No\,\,2, pp.\,241--258.
\end{minipage}\hfill
\\\vspace*{1.5mm}\\
\begin{minipage}[t]{86.5mm}
\noindent\hspace*{-5.2mm} %
Schulmeister, JC, Dahl, JM, Weymouth, GD and Triantafyllou, MS (2017). %
\lq\lq Flow control with rotating cylinders\rq\rq, {\it Journal of Fluid Mechanics},
Vol.\,825, pp.\,743--763.
\end{minipage}\hfill
\\\vspace*{1.5mm}\\
\begin{minipage}[t]{86.5mm}
\noindent\hspace*{-5.2mm} %
Schwaighofer, A, Tresp, V, Yu, K. (2005). %
\lq\lq Learning Gaussian process kernels via hierarchical Bayes\rq\rq, {\it Advances in neural information processing systems}, pp.\,1209--1216.
\end{minipage}\hfill
\\\vspace*{1.5mm}\\
\begin{minipage}[t]{86.5mm}
\noindent\hspace*{-5.2mm} %
Seeger, M (2004). %
\lq\lq Gaussian processes for machine learning\rq\rq, {\it International journal of neural systems},
Vol.\,14, No\,\,2, pp.\,69--106.
\end{minipage}\hfill
\\\vspace*{1.5mm}\\
\begin{minipage}[t]{86.5mm}
\noindent\hspace*{-5.2mm}
Sirisup, S, Karniadakis, GE, Saelim, N, Rockwell, D (2004). %
\lq\lq DNS and experiments of flow past a wired cylinder at low Reynolds number\rq\rq, {\it European Journal of Mechanics-B/Fluids}, Vol.\,23, No\,\,1, pp.\,181--188.
\end{minipage}\hfill
\\\vspace*{1.5mm}\\
\begin{minipage}[t]{86.5mm}
\noindent\hspace*{-5.2mm} %
Sun, S, Zhang, G, Wang, C, Zeng, W, Li, J, Grosse, R (2018). %
\lq\lq Differentiable compositional kernel learning for Gaussian processes\rq\rq, {\it arXiv preprint arXiv:1806.04326}.
\end{minipage}\hfill
\\\vspace*{1.5mm}\\
\setcounter{enumi}{\value{equation}}
\addtocounter{enumi}{1}
\setcounter{equation}{0}
\renewcommand{\theequation}{A.\arabic{equation}}

%


\end{document}